\documentclass{iaus}
\usepackage{graphicx}
\usepackage{graphics}
\usepackage{wrapfig}

\title[CO, HI, Recent Spitzer SAGE Results in the LMC] 
{CO, HI, Recent Spitzer SAGE Results in the Large Magellanic Cloud}

\author[Y. Fukui]   
{Yasuo Fukui}

\affiliation{Department of Physics, Nagoya University, Chikusa-ku, Nagoya 464-8602, Japan \break email: fukui@a.phys.nagoya-u.ac.jp}

\pubyear{2006}
\volume{237}  
\pagerange{119--126}
\date{?? and in revised form ??}
\jname{Triggered Star Formation in a Turbulent ISM}
\editors{B. G. Elmegreen \& J. Palous, eds.}
\begin{document}

\maketitle

\begin{abstract}
Formation of GMCs is one of the most crucial issues in galaxy
evolution. I will compare CO and HI in the LMC in 3 dimensional
space for the first time aiming at revealing the physical
connection between GMCs and associated HI gas at a $\sim 40$ pc
scale. The present major findings are 1) [total CO intensity]
$\propto$ [total HI intensity]$^{0.8}$ for the 110 GMCs, and 2)
the HI intensity tends to increase with the evolution of GMCs. I
argue that these findings are consistent with the growth of GMCs
via HI accretion over a time scale of a few $\times$ 10 Myrs. I
will also discuss the role of the background stellar gravity and
the dynamical compression by supershells in formation of GMCs.
\end{abstract}

\firstsection 
\section{Introduction}

The Magellanic system including the LMC, SMC, and the Bridge is an
ideal laboratory to study star formation and cloud evolution
because of its proximity to the sun. In particular, the LMC offers
the best site because of its unrivaled closeness and of the nearly
face-on view to us. Among the various objects in the LMC, the
molecular clouds which are probed best in the millimetric CO
emission, provide a key to understand star formation and galaxy
evolution. This is because the molecular clouds are able to
highlight the spots of star formation due to their highly clumped
distribution both in space and in velocity. The situation should
be contrasted with atomic hydrogen gas having lower density and
more loosely coupled to the star formation spots.

The key issue I would like to focus on in this talk is the
formation of giant molecular clouds (GMC) in the LMC. GMCs are the
major site of the star formation and the GMC formation must be a
crucial step in the evolution of a galaxy. In order to address
this issue, I will make (1) a comparison between HI and CO in 3
dimensional space to understand their physical correlation, (2) a
comparison between stellar distribution which is dynamically
controlling the HI density, and (3) a comparison with some of the
most recent Spitzer SAGE results. The contents of sections 3 and 4
are mainly based on the collaboration with Hinako Iritani and
Akiko Kawamura of Nagoya University and Tony Wong of ATNF and will
be published elsewhere.

\section{Three Classes of GMCs}\label{sec:GMC}

The NANTEN CO survey of the LMC has revealed that there are three
classes of GMCs according to the association of young objects
(Fukui et al.\ 1999; Yamaguchi et al.\ 2001b; Figure 1). Class I
has no apparent sign of star formation and Class II is associated
with small HII region(s) only but without stellar clusters.  Class
III is most actively forming stars as shown by huge HII regions
and young stellar clusters. This classification was presented by
Fukui et al. (1999) based on the first results of the NANTEN
survey (see also Mizuno et al.\ 2001; Yamaguchi et al.\ 2001b).
The basic scheme of the classification remains valid in the
subsequent sensitive survey with NANTEN while the number of GMCs
in each class has been increased by a factor of three (Fukui et
al.\ 2001; Fukui et al.\ 2006; see also Blitz et al. 2006). These
Classes are interpreted as indication of the evolutionary sequence
from I to III and the life time of a GMC is estimated to be a few
$\times$ 10 Myrs in total (Fukui et al.\ 1999). A comparison of
physical parameters indicates that size and mass tend to increase
from Class I to Class III, and Class III GMC has the largest size
and mass among the three. The stage after Class III is perhaps a
very violent dissipation of GMCs due to UV photons and stellar
winds from formed clusters as seen in the region of 30Dor most
spectacularly. More details on this classification are discussed
elsewhere in these proceedings (Kawamura et al.\ 2006).

\section{HI vs. CO}
\subsection{3-D correlation}

Previous studies of star formation in galaxies employed 2-D (2
dimensional) projection of HI intensity at large spatial averaging
over  $\sim$ 100 pc -- 1 kpc (e.g., Schmidt 1972 ). We shall here
test a 3-D (3 dimensional) comparison of CO and HI in the LMC
where the 3-D datacube has a velocity axis in addition to the two
axes in the sky. We use the 3-D datacube of CO with NANTEN (Fukui
et al. 2006) and of HI with ATCA (Kim et al.\ 2003), where CO
traces GMCs and HI less dense atomic gas.

Figure 2 shows an overlay of CO and HI and represents that GMCs
are located often towards HI filaments. This shows that the CO
emission is certainly located towards HI peaks while there are
also many HI peaks or filaments without CO emission,
\begin{wrapfigure}{l}[0pt]{63mm}
\begin{center}
\scalebox{0.33}{
 \includegraphics{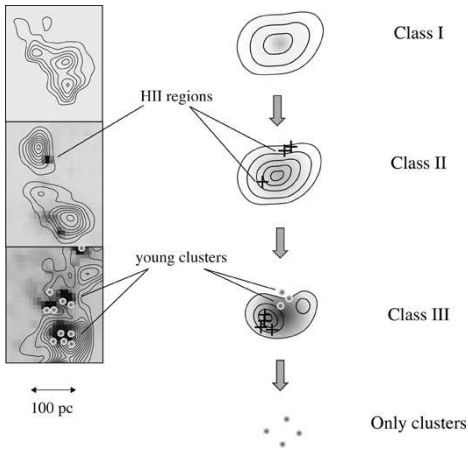}
 }
 \end{center}
\caption{
Evolutionary sequence of the GMCs in the LMC. An
example of the GMCs and illustration at each class are shown in the
left panels and the middle column, respectively. The images and
contours in the left panels are H$\alpha$ (Kim et al.\ 1999) and CO
integrated intensity by NANTEN (Fukui et al.\ 2006).
}
\end{wrapfigure}
suggesting that HI is the placental site of GMC formation (Blitz
et al.\ 2006). We should note that this overlay is of the
integrated intensities along the line of sight in velocity space.
Figure 2 shows typical CO and HI profiles in the LMC, indicating
that the CO emission is highly localized in velocity; the HI
emission ranges over 70 km s$^{-1}$ while the CO emission has a
width less than 10 km s$^{-1}$. The large velocity dispersion of
HI is most likely due to physically unrelated velocity components
in the line of sight and the HI gas associated with the CO gas is
a small fraction of HI whose velocity is close to that of CO.
Previous studies to correlate HI and CO using such velocity
integrated 2-D data obviously overestimates the HI intensity.

\begin{figure}
\centerline{
\scalebox{0.7}{
 \includegraphics{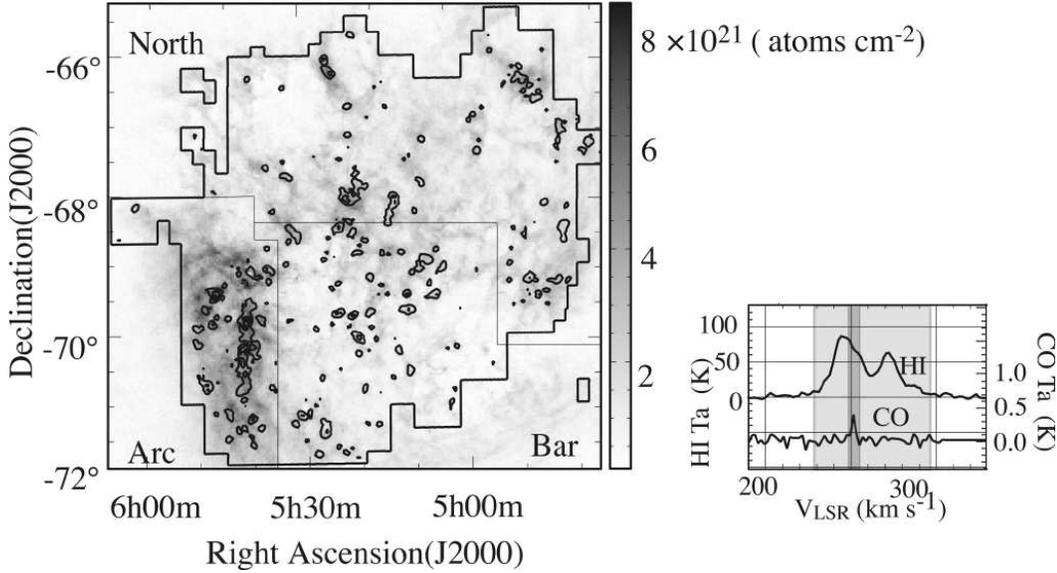}
 }
}
  \caption{{\it Left Panel}; HI image (Kim et al.\ 2003) with the CO contours (Fukui et al.\ 2006). The contours are from 1.2 K km s$^{-1}$ with 3.6 K km s$^{-1}$ intervals. The {\it Right Panel} shows an example of the HI and CO profiles at $\alpha$(J2000) $= 5^{h}35^{m}42^{s}$,
  and $\delta$(J2000) $= -69^{\circ}11'$.}
\end{figure}

The present 3-D analysis is able to pick up the HI gas physically
connected to the molecular clouds. Accordingly, we expect it
reveals the exact connection of the ambient atomic gas to the
dense molecular gas. The present HI and CO datasets have a pixel
size of 40 pc $\times$ 40 pc $\times$ 1.7 km s$^{-1}$ and consist
of $\sim$ 2million pixels. The HI and CO intensities are expressed
as averaged $T_{\rm a}$ (K) in a pixel with the lowest value, the
3$\sigma$ noise level, in each.

\begin{figure}[h]
\begin{minipage}{0.5\hsize}
\begin{center}
\includegraphics[width=60mm]{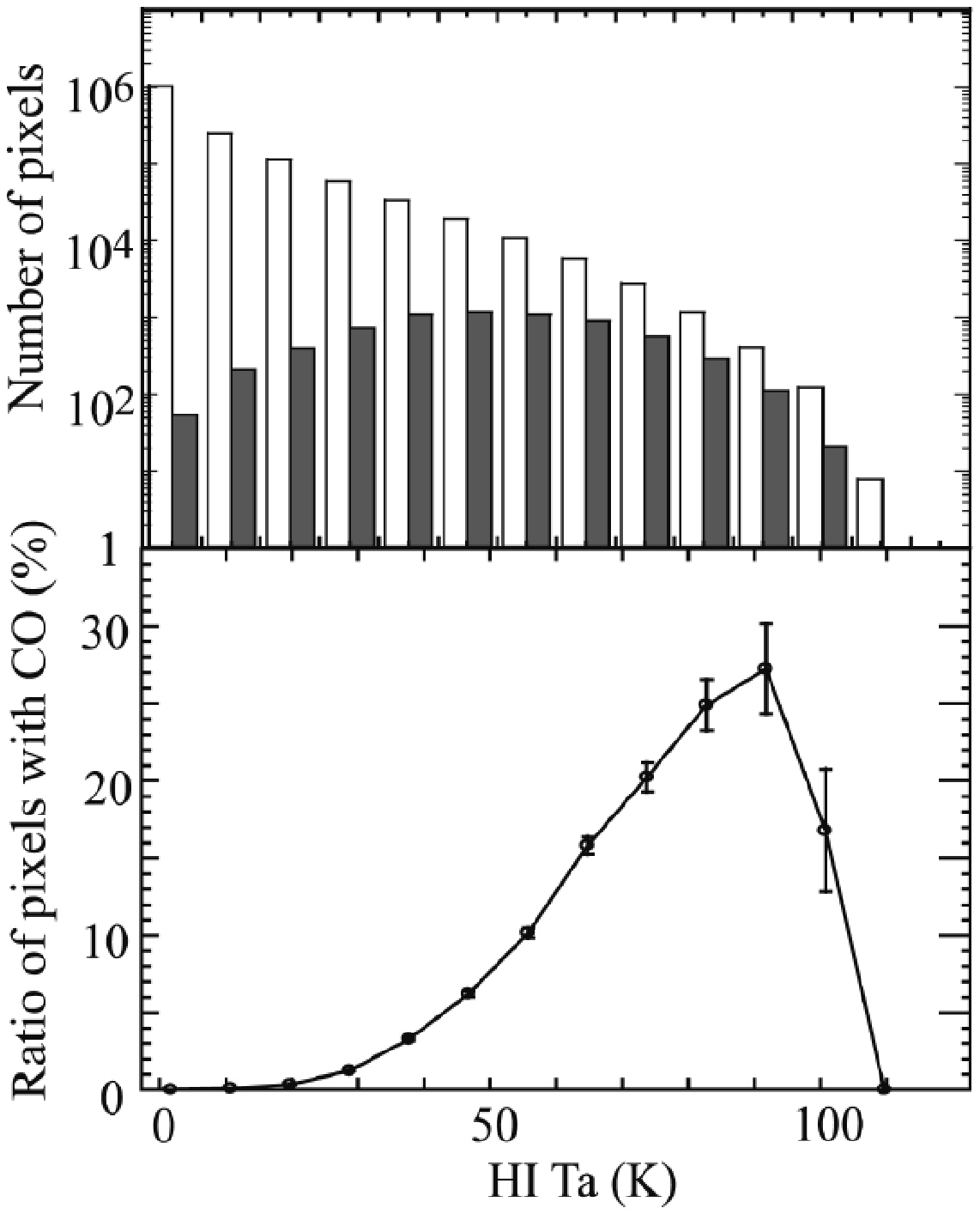}
\end{center}
\caption{
({\it Upper Panel}) Distribution of HI integrated intensity in 3-dimension.
Distribution of the HI integrated intensity where the significant CO emissions
are detected is also shown in gray.
({\it Lower Panel}) The ratio of the HI pixels with and without CO emission.}
\end{minipage}
\ \ \
\begin{minipage}{0.47\hsize}
\begin{center}
\includegraphics[width=67mm]{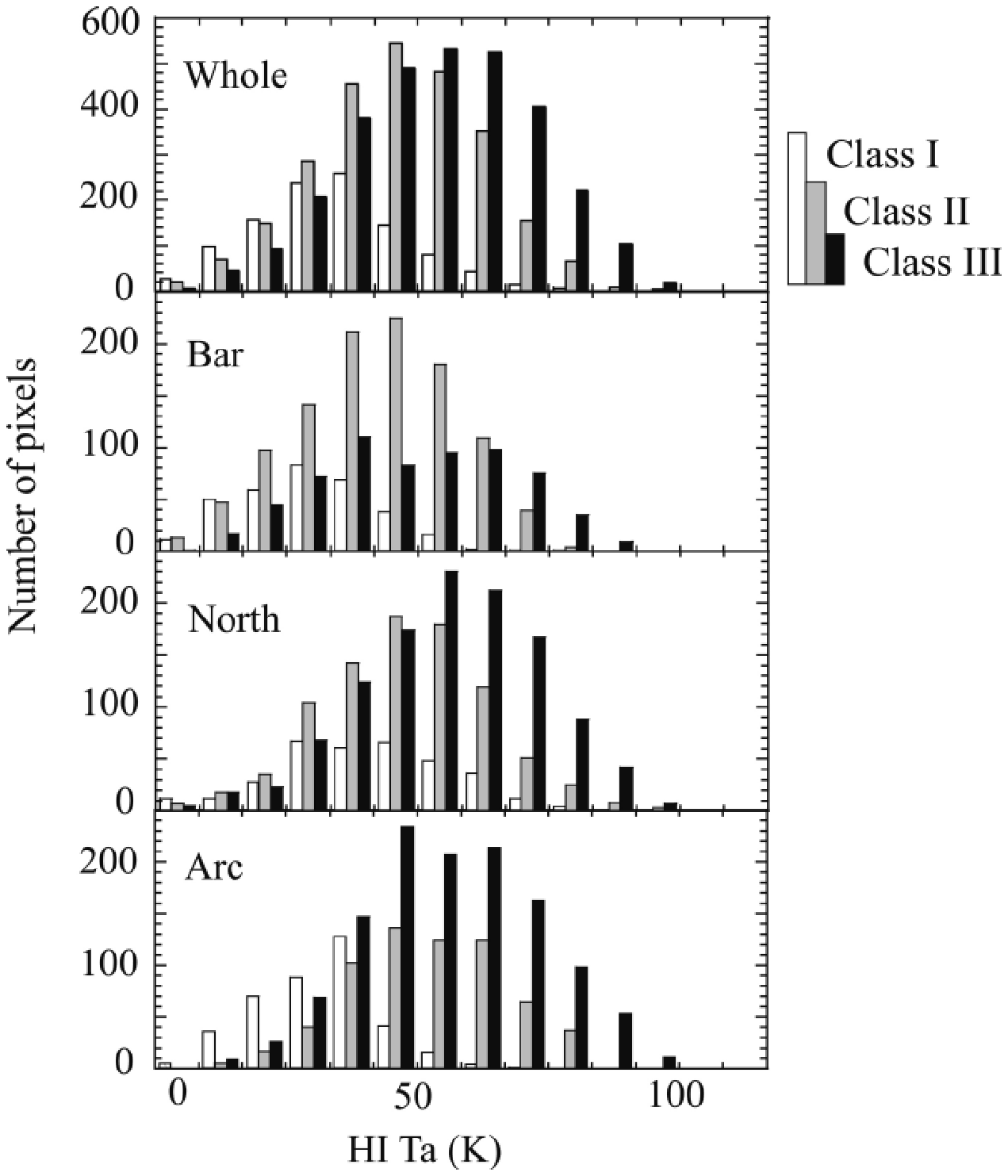}
\end{center}
\caption{
Distribution of the HI antenna temperature($T_{\rm a}$)
associated with the GMCs in the entire galaxy,
the arc , the bar, and the north regions from the top.
Each panel indicates the HI intensity distribution associated with
the GMC Class I (white), Class II (gray), and Class III (black). }
\end{minipage}
\end{figure}

Figure 3 shows a histogram of the HI integrated intensity in 3-D
and the pixels with the significant CO emission (greater than 0.7
K km s$^{-1}$) are marked. This histogram shows that the CO
fraction increases steadily with the HI intensity, suggesting the
HI provides a necessary condition to form GMCs.  About one third
of the pixels exhibit CO emission near $T_{\rm a}$(HI) of $\sim
90$ K and it seems that there is no sharp threshold value for GMC
formation with respect to the HI intensity.

\begin{figure}
\centerline{
\scalebox{0.6}{
\includegraphics{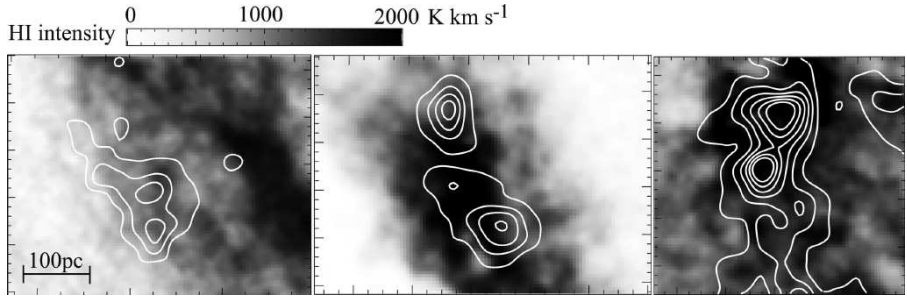}
} } \caption{Examples of the HI and CO distributions of the GMC
Class I({\it Left}), II({\it Middle}), and III({\it Right}),
respectively. Gray images are HI integrated intensity maps;
velocity is integrated over the range where the significant CO
emissions are detected The contours are CO integrated intensities
from 1.2 K km s$^{-1}$ with 1.2 K km s$^{-1}$ intervals(Fukui et
al.\ 2006).}
\end{figure}

Figure 4 shows a histogram of the HI intensity for the three GMC
Classes. This clearly shows that the HI intensity tends to
increase from Class I to Class III while the dispersion is not so
small. The average HI intensity over the whole LMC is 34 $\pm$ 16
K, 47 $\pm$ 17 K and 56 $\pm$ 19 K for Class I, II and III,
respectively. The HI intensity surrounding GMCs becomes greater
with the GMC evolution and star formation. In order to test the
variation within the galaxy, we shall divide the galaxy into three
regions, i.e., Bar, North and Arc (see the left panel of Figure
2). Histograms for each shown in the lower three panels of Figure
4 again indicate the same trend.

 \begin{wrapfigure}{l}[0pt]{78mm}
\begin{center}
\scalebox{0.35}{
 \includegraphics{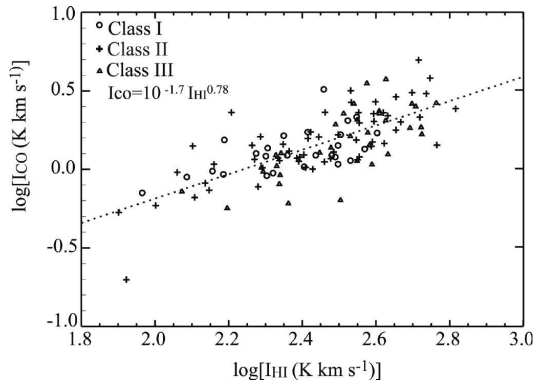}
 }
 \end{center}
\caption{
Correlation diagram between HI and CO intensities.
 Open circles, crosses and open triangles represent GMC Class I,
 Class II, and Class III, respectively. The dotted line shows the
 regression line of $I_{\rm CO} = 10^{-1.7} I_{\rm HI}^{0.78}$.
}
\end{wrapfigure}

Three images in Figure 5 show the HI distribution associated with
GMCs of the three Classes. The CO distribution has detailed
structures of $\sim 100$ pc or less and the HI seems to be
associated with the GMC at a scale of 50--100 pc.  The HI is not
always isotropic with respect to a GMC but indicates close spatial
correlations so that the HI is more or less enveloping a GMC.
 The typical HI velocity width associated is estimated to be $\sim$ 10--14 km s$^{-1}$,
and beyond this velocity span the associated HI peaks generally
becomes much weaker or disappear.

We have chosen 110 GMCs with five or more pixels in the two axes
in the sky and estimated $L_{\rm CO}$ in the velocity range where
CO is detected. Then we divide the $L_{\rm CO}$ by the apparent
size of a GMC. This gives an average CO intensity. For each GMC we
chose the pixels where the CO emission is significantly detected
and tested HI and CO correlation (Figure 6), where only the HI
pixels with CO are counted.  The regression shown in Figure 6 is
well fitted by a power law with a negative index of $\sim$0.8,
indicating a nearly linear correlation between CO and HI in a GMC.
We shall note that this index should become larger as $\sim$1.5 if
we use 2-D correlation because the velocity integrated HI
intensity has large offsets unrelated with CO.

\subsection{Dressed GMCs; Growth of GMCs via HI Accretion}

The dependence of HI intensity on Class of GMC indicates that the
surroundings of a GMC change appreciably depending on the Class at
a time scale of 40--100pc. The HI intensity is generally a product
of the spin temperature and the optical depth, and the correlation
indicates temperature and/or density is dependent on Class. The
spin temperature of HI is generally estimated to be $\sim$ 100 K
in the Milky Way. Since the HI spin temperature may be higher in
the LMC where UV is more intense and dust opacity is less than in
the Milky Way. The typical HI intensity less than 100 K suggests
that the HI emission is optically thin. If so, the HI intensity
should represent HI density. We therefore infer that GMCs are
``dressed" in HI and that the ``HI dress" grows in time.

The correlation between HI and CO is nearly linear (Figure 6).
This alone does not provide a strong constraint on the formation
of a GMC. Nonetheless, the apparent association of HI with GMCs
suggests that the HI is enveloping each GMC and the HI density
increases with the cloud evolution.  We infer that this represents
the enveloping HI gas is accreting onto GMCs and is converted into
H$_{2}$ due to increased optical depth. This leads to increase the
molecular mass of GMC, i.e., the observed mass increase from Class
I to III (see section 2). The timescale of the GMC evolution is
$\sim$ 10 Myrs and the increased molecular mass is in the order of
$10^{6} M_{\odot}$. Namely, a mass accretion rate of$\sim 10^{-1}
M_{\odot}$ yr$^{-1}$ is required. We roughly estimate that this
rate is consistent with that calculated for a spherical accretion
of the HI gas having $n$(HI) $\sim 10$ cm$^{-3}$ at an infall
speed of $\sim$ 7 km s$^{-1}$.

\section{Stellar Gravity and Triggering in GMC Formation}

We shall examine two effects which may be important in converting
HI to H$_{2}$: one is the stellar gravity and the other
supershells driven by OB stars and/or SNRs.

\subsection{Stellar gravity}

We shall here use the K-band image of 2MASS which represents
relatively old stellar population dominating the gravity (Sergei
et al.\ 2000). This will allow us to test the effects of stellar
gravity on GMC formation.  The young stars associated with GMCs
are small in mass and are not important in such gravity.

Figure 7 shows a histogram of CO clouds vs. K-band stellar
density. We find that the number density of CO clouds increases
with the increase of the stellar density. It seems that there is a
threshold value at $\sim 1.0$ in log (number of stars/4
arcmin$^{2}$) for the HI - H$_{2}$ conversion where the number
density of CO clouds increases beyond $\sim 5$ \% of the pixel
number.

Wong and Blitz (2002) argue that the pressure exerted on HI gas
may play a role in converting HI to H$_{2}$. They studied several
spiral galaxies to show that the HI gas is associated with CO only
in the inner part of galaxies where the stellar gravity is strong.
The pressure should be basically dominated by the stellar gravity
as long as the stars are the dominant source of gravity and the
present result on the LMC supports that the pressure plays a role
in converting HI into H$_{2}$.

\clearpage

\subsection{SAGE Results}

\begin{wrapfigure}{l}[0pt]{75mm}
\begin{center}
\scalebox{0.54}{
 \includegraphics{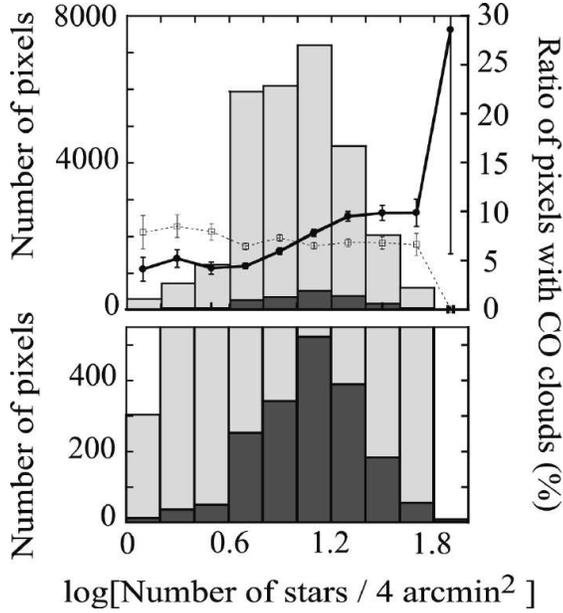}
 }
 \end{center}
\caption{ Distribution of the stellar density derived from the
2MASS K band sources.   The histogram in dark gray shows the
stellar density distribution where the significant CO emissions
are detected. Thick line indicates the ratio of the pixels of the
stellar density with and without CO emissions. Thin line shows the
ratio as for the thick line but obtained by assuming the CO
emissions are distributed randomly. The {\it Lower Panel} is an
enlargement of the  {\it Upper Panel}. }
\end{wrapfigure}

The Spitzer satellite has been used to make an extensive study of
the LMC at infrared wavelengths from 3.6--160$\mu$m.  This is the
SAGE project headed by M. Meixner.  Figure 8 shows the 3.6--24
$\mu$m 3-color composite image (3.6$\mu$m, 8.0$\mu$m and 24$\mu$m
in blue, gree, and red) from the SAGE project
(NASA/Caltech-JPL/Meixner STScI and the SAGE Legacy Team) with
NANTEN CO distribution (Fukui et al.\ 2006). This presents that CO
clouds are well correlated the far-infrared distribution and
future comparisons should reveal considerable details of  the
dust-gas relationship quantitatively.

In connection with the GMC formation, we note that the SAGE image
indicates numerous good candidates for apparently swept-up matter
perhaps by supershells owing to the better angular resolution.
Yamaguchi et al. (2001a) made a comparison between H$\alpha$
supershells and CO and concluded that $\sim 30\%$ of the GMCs may
be associated with supergiant shells. Since more candidates may be
identified from the SAGE image, triggering of GMC formation by
supershells may be more important than previously thought by
Yamaguchi et al.\ (2001a).

\section{Summary}

I shall summarize this contribution as follows;\\
1) 3-D comparison between CO and HI has revealed GMCs are associated with HI gas at a $\sim$ 40pc scale; these are ``HI dressed GMCs".\\
2) There is a clear increase of the HI intensity around GMCs from GMC Class I to Class III. Growth of GMCs in mass via HI accretion has been suggested over a time scale of a few  $\times$ 10 Myrs.\\
3) This correlation has a form, [CO intensity] $\propto$ [HI intensity]$^{0.8}$, for selectec 110 major GMCs.\\
4) The background stellar gravity and dynamical compression by supershells may be important in converting HI into H$_{2}$, i.e. in GMC formation.

\begin{figure}
\centerline{
\scalebox{0.8}{
 \includegraphics{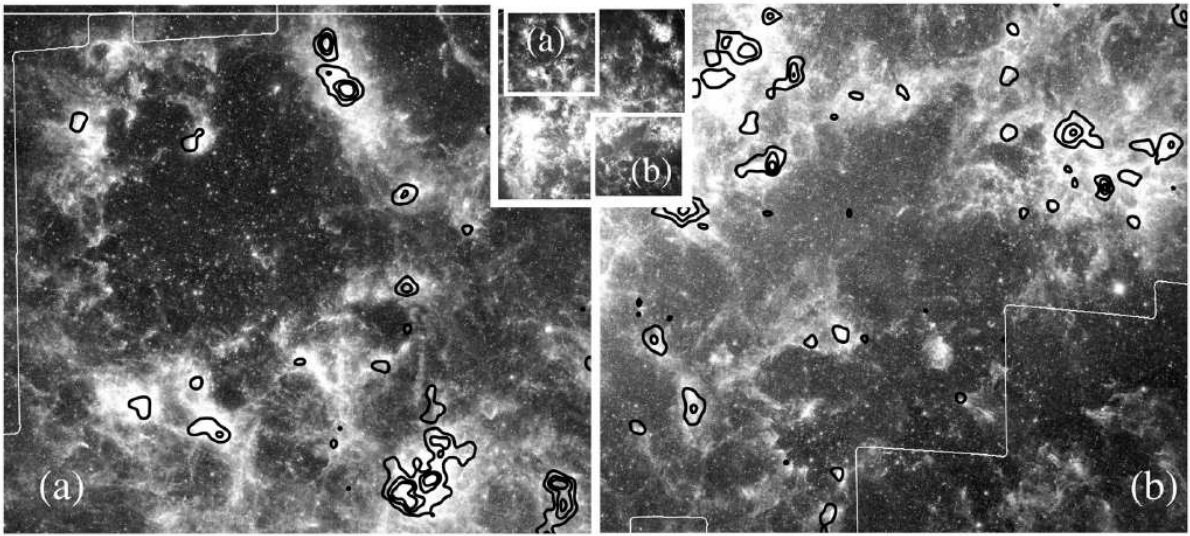}
 }
}
  \caption{3.6--24 $\mu$m 3-color composite image (3.6$\mu$m, 8.0$\mu$m and 24$\mu$m in blue, gree, and red, respectively) from the SAGE project (NASA/Caltech-JPL/Meixner STScI and the SAGE Legacy Team) of the LMC ({\it Middle} inset);
  Overlays with CO contours are shown
(a) toward LMC 4, and (b) LMC 7 and 8.
  The CO contours are from 1.2 K km s$^{-1}$ with 2.4 K km s$^{-1}$  intervals
  from the CO survey by NANTEN (Fukui et al.\ 2006).
  The regions of (a) and (b)
  are indicated by white boxes in the inset.}
\end{figure}

\begin{acknowledgments}
The NANTEN project is based on a mutual agreement between Nagoya
University and the Carnegie Institution of Washington. We greatly
appreciate the hospitality of all the staff members of the Las
Campanas Observatory. We, NANTEN team, are thankful to many
Japanese public donors and companies who contributed to the
realization of the project.  We would like to acknowledge Drs. L.
Staveley-Smith and S. Kim for the kind use of their HI data. We
are also thankful to IRAC and MIPS pipeline team to create a
beautiful images of the LMC by Spitzer. This work is financially
supported in part by a Grant-in-Aid for Scientific Research from
the Ministry of Education, Culture, Sports, Science and Technology
of Japan (No.\ 15071203) and from JSPS (No.\ 14102003,
core-to-core program 17004 and No.\ 18684003).
\end{acknowledgments}

\end{document}